# Analytical stochastic macroscopic fundamental diagram driven by Wiener process

HongSheng Qi

([1] College of Civil Engineering and Architecture, Zhejiang University. 866 Yuhangtang Road, Hangzhou, 310058, China

**Abstract**: The macroscopic fundamental diagram (MFD) is a powerful and popular tool that describes a network scale traffic operational state and serve as the plant model of perimeter control. As both the supply and the demand suffer from random disturbances, the traffic flow dynamics cannot be said to be deterministic. A stochastic MFD model which can generate a stochastic evolution of the system state with desired distribution of aggregated variables is still lacking. A stochastic formulation of MFD with explainable parameters which can be calibrated using observations is proposed to fill this gap. The model is based on the stochastic differential equation (SDE) theory. First, the exit flow variation is formulated as a Wiener-driven process, which admits the accumulation-dependent variations. The stochastic MFD model is then constructed by combining the exit flow variations model. The solution of the system state is derived by the forward Fokker-Planck equation. The stability of the model is analyzed, and the parameters of a calibration method are provided. Several cases of the model are then tested. The results show that the model can be applied to different functional MFD forms, and the hysteresis and gridlock phenomenon is reproduced. The proposed MFD model can be used in the network analysis and control that considers the system's stochastic evolution.

**Highlights:**

- MFD's stochasticity is investigated by a stochastic differential equation theory with explainable noise parameter;
- The state-dependent exit flow variation is described by a SDE model driven by the Wiener process;
- The model is independent of a specific functional form of MFD. Its analytical solution, stability condition, and parameter calibration method are provided.
- Case study shows that the model can reproduce the gridlock process, hysteresis phenomena. Marginal distributions can be obtained also.

# 1  Introduction

## 1.1  Background

Traffic flow control and management rely on the operational analysis of the network. While there are many methods that can perform such tasks, the MFD approach receives more and more attention. Since the empirical proof of its existence in 2007, frameworks of MFD have become more and more systematic, including theoretical model development, property analysis, calibration, application, etc. According to Johari et al. (2021), the MFD can now be applied to a variety of traffic flow scenarios (Simoni, Waraich, and Hoogendoorn, 2015).

The MFD should be stochastic, rather than deterministic, as it tries to describe the network-level traffic flow operational state. That state depends on two meta-causes: supply and demand. Both sides suffer from random factors. For instance, the factors that influence the capacity side include its random network capacity (Jin et al. 2009); traffic incidents; unscheduled maintenance, etc. The factors that influence the demand side include: random travel behavior (departure time, route choice, etc.) as well as daily or seasonal demand fluctuations, etc. The network traffic flow -level, which is the result of interaction between the demand and supply, cannot be said to be deterministic. Therefore, there is a need to investigate the MFD from a stochastic viewpoint. Two examples of the stochastic nature of the MFD are presented in Figure 1. Figures 1-(a) and (c) are the case for simulated MFD using a network of the size 10*10 in SUMO, and Figures 1-(b) and (d) are real-world cases, which is calculated by Edie definition (Saberi et al. 2014). The MFD shows a general trend, similar to those found and demonstrated in much of the literature. When given the same density, production inevitably presents fluctuations. Of more importance, the distribution pattern of production is density-dependent. Given that one network density corresponds to many production values, the network experiences different paths during the stages of loading or unloading. Further, daily or seasonal evolution paths do not repeat themselves. This is demonstrated in Figure 2, which gives the network flow rate and density (aggregated by Edie definition by Saberi et al., 2014) for 7 continuous days using real world data. The time axis is 24 hours. It is shown that,although a certain pattern of the system evolution exists, the fluctuation cannot be neglected. Evidently, the deterministic approach is not enough.

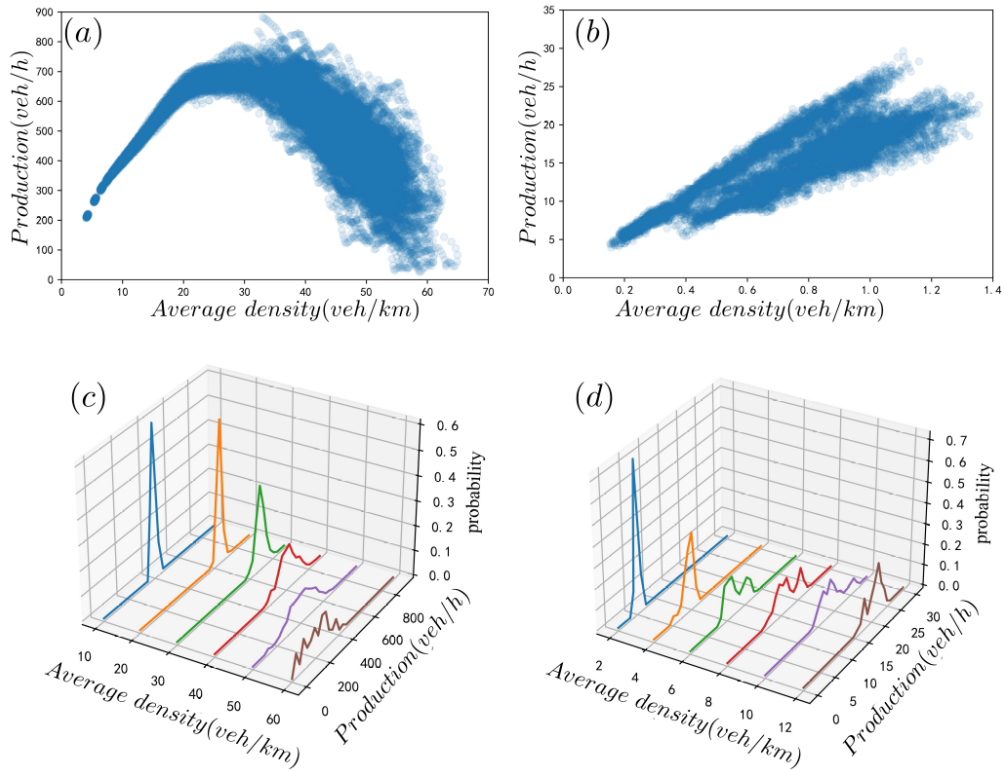

Figure 1. The distribution of simulation (left) and real world data.

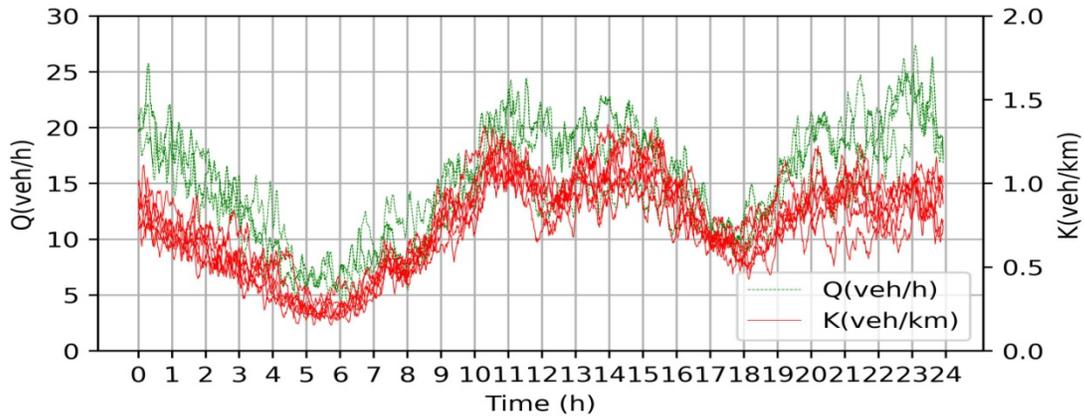

Figure 2. The evolution of network flow rate and density (calculated from Edie definition) for 7 days.

Some research identifies and recognizes the wide distribution of network flow rate or production under the same density (Daganzo et al., 2011; Mahmassani et al., 2013). According to some analyses, the major reason is that the congestion is heterogeneous within the region (Jin et al. 2013; Gan et al., 2017). Three methods are available to deal with this issue (Johari et al.,2021). One way is to resort to the region division method, i.e., the network should be partitioned based on the traffic flow state (Ji and Geroliminis, 2012). The second is to develop a functional form of MFD (Mahmassani et al., 2013b). These two functional forms take the congestion heterogeneous as an independent variable of MFD. The third method is to fit the MFD for loading or unloading separately (Paipuri et al., 2019). None of the above three approaches addresses the stochastic evolutions of the system states. In Gao and Gayah (2017), a Markov chain framework was used to approximate the stochastic differential equation of the MFD, which is formed by adding a noise term in conservation equation, rather than the trip completion rate function. Therefore the model still admits that relationship between the aggretated varaibles is deterministic. An analytical approach that can produce the distributions with explainable noise is still lacking. There is much literature that has its focus on a stochastic fundamental diagram on roads (Gu et al., 2017; Siqueira et al., 2016; Ni, et al., 2018). However, the MFD of an analytical stochastic type is not mature,as compared with the fundamental diagram of a road. Moveover, the parameters in the model should be explainable and can be calibrated using observations. Existing literaters considering stochasticity include the approximation method (Laval and Castrillón, 2015), or the discretized form using Markov chain (Gao and Gayah, 2017).

In response to the above research gap, we investigated the MFD from a stochastic viewpoint, and assumed that there exists an upper and a lower boundary of the aggregation level variable. The system state (which can be described by selected aggregation level variables) fluctuates within the boundaries and is driven by some stochastic process, given the input demand. In this way, the distribution of the density/production, the evolution path, etc. can be formulated or derived.

## 1.2 Literature Review

Using the aggregate level variables to represent the traffic flow state is not a new idea and can be dated back to 1969 (Godfrey, 1969). The empirical observation proves it in 2008 (Geroliminis and Daganzo, 2008). Since then, the MFD has received more and more attention, and is applied in many tasks, including perimeter control, congestion charging, routing, stabilization (Sirmatel and Geroliminis, 2021; Ding et al, 2020), etc. The physical meaning or the interpretation of MFD can be classified as the following two types (Ambuhl et al. 2021). The first type views the MFD as a theoretical upper boundary of the performance (network flow rate) and is independent of the true demand; while the second type uses real world data to generate the MFD shape, and can be interpreted as the observed MFD, which should be lower than the first type (Mahmassani et al. 2013).

Under the umbrella of the above two interpretations, there are many formulations of the MFD. One most commonly used formulation (Type-I) is the average network flow and average density. The variables are defined as follows: average flow (or production) is calculated as the weighted sum of the flow rate on all or partial roads. The average network density is the weighted sum of all or partial roads. The second type (Type-II) uses the total traveled distance (TTD), instead of production, and total traveled time spent (TTT), instead of density. These variables can be calculated from trajectory data (Keyvan-Ekbatani et al. 2019). The third type (Type-III), that is similar to the first type, uses the trip completion flow, rather than the production (Sirmatel and Geroliminis 2021; Zhong et al., 2017) to represent the y-axis variable. This type is very like the original formulation proposed by Daganzo (2007), where the trip completion rate is named "exit flow" which describes the assumed single reservoirs. When the average trip length is a constant, there is a linear relationship between the trip completion flow and the production, or network average flow (Ampountolas, Zheng, and Geroliminis 2017). The functional type of the MFD varies across different literature. For instance, there are polynomials (Sirmatel and Geroliminis 2021; Zhong et al. 2017; Ding et al. 2017), logistic-based models; smooth approximation of a trapezoidal model (Ambühl et al., 2017); exponential family (Geroliminis et al. 2014), etc.

Currently, most MFD models are developed mainly as a deterministic system (Wan et al. 2020). As indicated in Johari et al. (2021),"NMFD relations do not recognize stochasticity per se". The stochastic counterpart is still limited. However, many real-world observations reveal a certain scattering of network flow and density (Ding et al, 2020). Moreover, the scattering or spatial heterogeneity depends on the accumulation or network density. The spatial heterogeneity can be described by the standard deviation (Mahmassani et al. 2013; Muhlichet al. 2015) or variance (Ji and Geroliminis, 2012). To deal with randomness contained in the scattering, three technical approaches can be found in the current literature: 1) some noise terms are added to the deterministic MFD formulation (Zhong et al., 2017; Haddad and Mirkin, 2020; Hajiahmadi et al., 2015); 2) uncertain boundaries are considered and robust measures can be designed (Ampountolas, Zheng, and Geroliminis, 2017; Haddad and Shraiber, 2014); 3) certain distributions of the accumulation can be fitted (Ding et al, 2020). Haddad and Mirkin (2020) include a noise term in the exit flow function. This noise is treated as a uniform distribution in a pre-specified interval. Haddad and Shraiber (2014) assume that the uncertain MFD is a function that lies between the upper and lower boundaries of specified MFD, and is designed as a robust method for perimeter control. Ding et al (2020) admitted the fluctuation of the exit flow, and found that the distribution of that exit is uneven between the upper and lower boundary exit flow. Hajiahmadi et al. (2015) added an error term, that follows uniform distribution, to account for the exit flow variation.

The above MFD formulations provide a plethora of tools to model the network traffic flow. As both sides of network traffic (demand and supply) suffer from random factors, it is natural to equip the traditional MFD with stochastic dynamics. In Gao and Gayah (2017), the stochastic traffic flow dynamics was modeling by adding noise term in the flow conservation equation. The employed trip completion function still is deterministc; In Laval and Castrillón (2015), an probabilistic method which approximates the MFD at a corridor is formulated. To develop an analytical stochastic MFD, the accumulation dependent variables need to be considered with explainable noise parameters. Unfortunately, the stochastic framework of MFD is not so mature, as compared with the deterministic counterpart.

## 1.3 Summary, scope and manuscript organization

The stochasticity in network level traffic flow operation behaves differently under different density levels. Due to random factors, the system path evolves in a stochastic way. To the authors' knowledge, a MFD model with explainable noise parameters that can generate distributions of the desired aggregated-level variables is still lacking. In response to the above needs, we propose a stochastic MFD model, which is based on a stochastic differential equation theory. The model is derived from a transformed Brownian motion and admits that there are density-dependent exit flow variations. The parameter in the model has a clear physical meaning and can be calibrated by observations. The proposed model can produce distributions of the exit flow, under a given density, and can also generate the hysteresis and gridlock phenomenon.

The manuscript is organized as follows. We develop the model in Section 2. The solution to the model, which is represented by the Fokker-Planck equation, stability analysis, and the calibration method, is presented in Section 3. In Section 4, we validate our model by real-world data and simulation data, and conclude with some remarks in Section 5.

## 2 Model

### 2.1 Notations

$n_i$: the number of the vehicles at region i.

$n_{in,i}^{buf}(t)$: the number of vehicles in the buffer area. The buffer area stores the vehicles that fail to enter the region.

$q_{in,i}^{raw}(t)$: the raw inflow rate for region i;

$\Psi(x)$: the dummy function, which is defined as $\Psi(x) = \frac{x}{\sqrt{m+x^2}}$

$\Omega_{in,i}\left(n_{in,i}^{buf}(t), n_i(t)\right)$: the final input flow into region i;

$n_{i,jam}$: the vehicles numbers that correspond to the jam state.

$q_{ij}(t)$: the flow from region i to region j;

$q_{MAX}(t)$: the maximal input demand into some region.

$G(n)$: the exit flow rate, which is a stochastic process;

$Z(n)$: the accumulation-dependent exit flow variation, which is a stochastic process driven by the Wiener process.

$g_{mi}(n_i(t))$: the expected outflow function of the region. It can vary for different regions;

$g_{lw}(n_i(t))$ and $g_{up}(n_i(t))$: the upper and lower boundaries of the exit flow;

$\Gamma^+(n)$ and $\Gamma^-(n)$ are defined as $\Gamma^+(n) = g_{up}(n) - g_{mi}(n)$, $\Gamma^-(n) = g_{mi}(n) - g_{lw}(n)$.

$\Delta^-(K)$ and $\Delta^+(K)$: are defined as $\Delta^-(K) = \Gamma^+(K) - \Gamma^-(K)$ and $\Delta^+(K) = \Gamma^+(K) + \Gamma^-(K)$;

dB: Brownian motion;

σ: the parameters in the MFD

## 2.2 Model development procedure

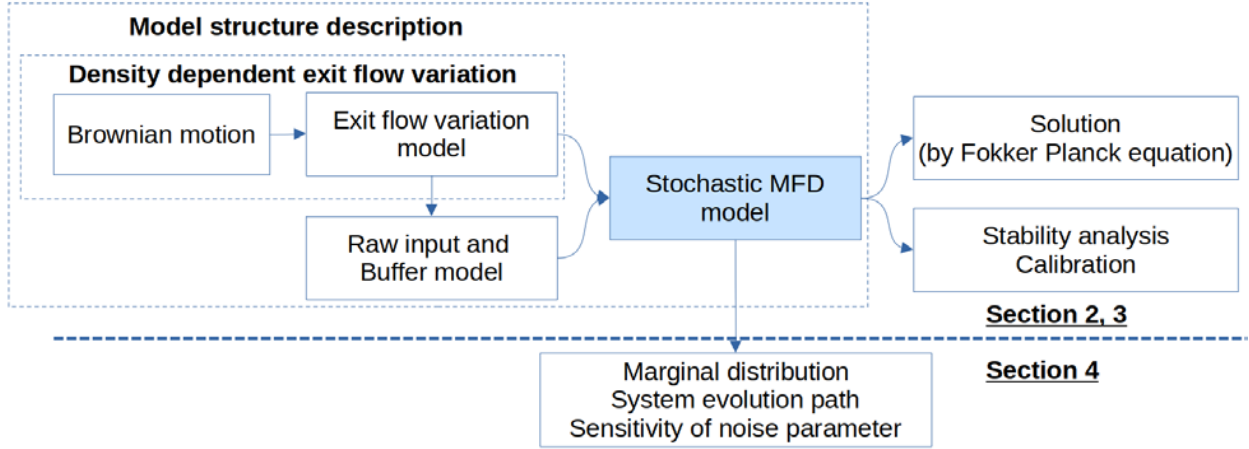

Figure 3. Procedures for model development.

The development of our model follows the procedures illustrated in Figure 3. We first describe the model structure in the remaining paragraph of this section. The model admits the common accumulation-based exit-MFD formulation while suggesting a certain exit flow variation (developed in Section 2.4). The exit flow variation model is constructed from a basic Brownian motion. After that, a buffer that stores the vehicles that fail to enter the region, due to congestion, is taken into account. Together, with exit flow variation, the stochastic MFD is derived. The solution of the stochastic MFD model, which is represented by the Fokker-Planck equation, is derived, and a calibration method is proposed. In Section 4, we apply our model to several cases, including a real-world case and simulated data.

## 2.3 Overall model structure

We do not assume a single map of the trip completion rate from an accumulation of the number of vehicles, in the region, to the exit flow. Rather, the trip completion rate is distributed within a certain range under the same accumulation. We assume that there is a certain expectation relationship between the accumulation and the exit flow rate. Such relaship can be obtained easily obtained by averaing the exit flow rate for certain accumulation. Moreover, as we can seen in the experiment in section 4.4, the relationship maynot be necessary the expectation of the exit flow observations. Also for a region, the upper and lower bound of the exit flow under the same accumulation can be determined with real world observation. The upper boundary, lower boundary, and expectation are indicated by $g_{up}(n)$, $g_{lw}(n)$ and $g_{mi}(n)$ respectively, as in Figure 4, where 'mi' represents the middle. The observed relationship is scattered in the upper and lower boundary areas. The difference between the three curves is expressed as $\Gamma^+(n) = g_{up}(n) - g_{mi}(n)$, $\Gamma^-(n) = g_{lw}(n) - g_{mi}(n)$. Note that, as we aforementioned, the expectation $g_{mi}(n)$ may not necessarily be the middle point between the upper and lower boundaries. It can be calibrated, or adjusted, as required. Besides, the three characteristic curves of the macroscopic level variables can be specified in a flexible way for each region.

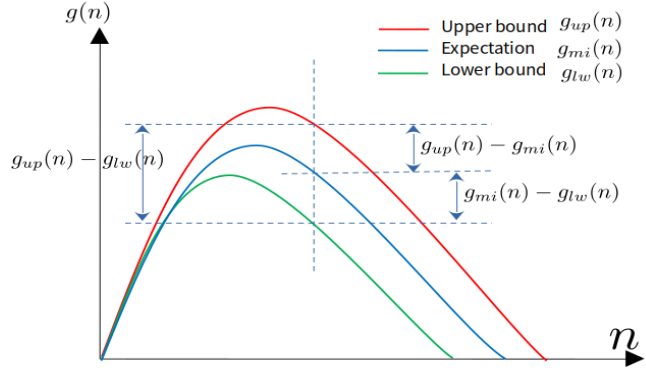

Figure 4. The formulation of the exit function.

For single region i, the accumulation is the summation over all possible paths, i.e., $n_i(t) = \sum_j n_{ij}(t)$, where $n_{ij}(t)$ is the vehicle's number from Region i to Region j. $n_{ii}(t)$ represent the endogenous vehicles generated within the region and complete their trip within the region. The conservation law for a single region i reads:

$$\frac{dn_i}{dt} = q_{in,ii}(t) - q_{out,ii}(t) + \sum_{k \neq i} q_{ki}(t) - \sum_{k \neq i} q_{ik}(t) \qquad \text{Eq. 1}$$

$q_{in,ii}(t)$ is the endogenous entrance flow rate, and $q_{out,ii}(t)$ is the exit flow rate. $q_{ki}(t)$ that represents the entering flow from other regions, and $q_{ik}(t)$ is the exiting flow to other regions. The outflow is proportional to the number of vehicles in this region, i.e.:

$$\begin{cases} q_{out,ii}(t) = \dfrac{n_{ii}(t)}{n_i(t)} G(n_i(t)); \\ q_{ij}(t) = \dfrac{n_{ij}(t)}{n_i(t)} G(n_i(t)), \forall j \end{cases} \qquad \text{Eq. 2}$$

$G(n_i(t))$ is the **realized** total trip completion rate at Region i. Clearly $G(n_i(t))$ is within $[g_{lw}(n), g_{up}(n)]$. We assume that $G(n_i(t))$ can be decomposed into the expected outflow plus a variation $Z_i(n_i(t))$:

$$G(n_i(t)) = g_{mi}(n_i(t)) + Z_i(n_i(t)) \qquad \text{Eq. 3}$$

And, hence, the conservation law in Eq. 1 becomes:

$$\frac{dn_i}{dt} = q_{in,ii}(t) - \frac{n_{ii}(t)}{n_i(t)} G(n_i(t)) + \sum_{k \neq i} \frac{n_{ki}(t)}{n_i(t)} G(n_k(t))$$

$$- \sum_{k \neq i} \frac{n_{ik}(t)}{n_i(t)} G(n_i(t))$$

$$= q_{in,ii}(t) - \left(\frac{n_{ii}(t)}{n_i(t)} + \sum_{k \neq i} \frac{n_{ik}(t)}{n_i(t)}\right)\left(g_{mi}(n_i(t)) + Z(n_i(t))\right)$$

$$+ \sum_{k \neq i} \frac{n_{ki}(t)}{n_i(t)} \left(g_{mi}(n_k(t)) + Z_k(n_k(t))\right)$$

Eq. 4

The above formulation is general, which allows different formulations of $g_{mi}(n_k(t))$ and $Z_k(n_k(t))$ for different regions. In the next two sections, we develop the model for $Z_k(n_k(t))$, and then complete the formulation of stochastic MFD in Eq. 4. $Z_k(n_k(t))$ is named density-dependent exit flow variation.

## 2.4 Density dependent exit flow variation formulation

The variation, i.e., $Z_k(n_k(t))$, can be interpreted as the stochastic operational noise. The source of noise is manifold. We interpret the noise in the following ways: the demand is stochastic, in that the temporal demand level fluctuates; the route choice is stochastic etc.; the supply is also stochastic (for instance the intersection capacity obeys a certain distribution); a random traffic flow incident etc. To develop the noise, we first assume that there is a meta-noise W, which is characterized by a classic Brownian motion (i.e., Wiener process):

$$dW = \sigma \cdot dB \qquad \text{Eq. 5}$$

B is standard Brownian motion and $\sigma$ is the diffusion factor, which can be calibrated. $\sigma$ can be interpreted as the stochasticity level. The range of W is $[-\infty, \infty]$. However, the variation, i.e., $Z_k(n_k(t))$ is within the interval of $[\Gamma^-(n_k(t)), \Gamma^+(n_k(t))]$. A transformation from a general stochastic process W to the exit flow variation is necessary. We select the hyperbolic tangent function to represent this transformation from W to Z (we omit the subscript, i.e., region k here):

$$Z = \Gamma^+(n) + \frac{\tanh(W) + 1}{2}\left(\Gamma^+(n) - \Gamma^-(n)\right) \qquad \text{Eq. 6}$$

And it follows that:

$$\tanh(W) = \frac{2(Z - \Gamma^-(n))}{(\Gamma^+(n) - \Gamma^-(n))} - 1 \qquad \text{Eq. 7}$$

As W is a stochastic process that is described in Eq. 5, then Z, which is a function of W, can be formulated using Ito theorem. To formulate the expression of Z, we first get the derivates of Z:

$$\begin{cases} Z' = \frac{(1-\tanh^2(W))}{2}(\Gamma^+(n) - \Gamma^-(n)) \\ Z'' = -\tanh(W)(1-\tanh^2(W))(\Gamma^+(n) - \Gamma^-(n)) \end{cases} \quad \text{Eq. 8}$$

Using the Ito theorem, we have the following expression of Z:

$$dZ = \frac{1}{2} \cdot \sigma^2 \cdot Z'' dt + \sigma \cdot Z' dB \quad \text{Eq. 9}$$

$$= -\frac{1}{2} \cdot \sigma^2 \cdot \left((\tanh(W) - \tanh^3(W))(\Gamma^+(n) - \Gamma^-(n))\right) dt + \sigma$$

$$\cdot \frac{(1-\tanh^2(W))}{2}(\Gamma^+(n) - \Gamma^-(n)) dB$$

$$= -\frac{\sigma^2 \cdot (-\Delta^+(n) + 2Z + (\Delta^+(n) - 2Z)^3)}{2} dt$$

$$+ \sigma \frac{\left(\Delta^-(n)\right)^2 - (\Delta^+(n) - 2Z)^2}{2(\Delta^-(n))} dB$$

In the second derivation, we use the fact that $\tanh(K) = \frac{2(Z-\Gamma^-(n))}{(\Gamma^+(n)-\Gamma^-(n))} - 1$. The noise Z depends on the parameter $\sigma$ and accumulation n.

## 2.5 SDE-MFD formulation

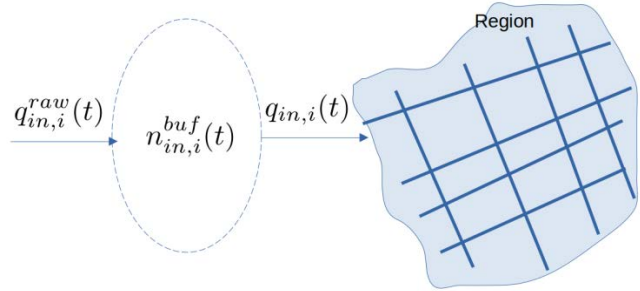

Figure 5. Illustration of the buffer.

We assume that there is a buffer area for each region. The system illustration is given in Figure 5. When the region is congested enough, then vehicles cannot successfully enter the region. These vehicles, that fail to enter, are assumed to be stored in a buffer area. The number of vehicles in the buffer is indicated by $n_{in,i}^{buf}(t)$. The true input to the region is $q_{in,ii}(t)$. When $n_{in,i}^{buf}(t) = 0$, the entering flow rate $q_{in,ii}(t)$ equals $q_{in,i}^{raw}(t)$; when $n_{in,i}^{buf}(t) > 0$, the instant entering flow rate is $q_{MAX}(t)$, which is a fixed parameter. The system equation of single Region i is expressed as Eq. 10:

$$\begin{cases} \frac{dn_i}{dt} = q_{in,ii}(t) - \left(\frac{n_{ii}(t)}{n_i(t)} + \sum_{k \neq i} \frac{n_{ik}(t)}{n_i(t)}\right)\left(g_{mi}(n_i(t)) + Z_i(n_i(t))\right) + \sum_{k \neq i} \frac{n_{ki}(t)}{n_i(t)}\left(g_{mi}(n_k(t)) + Z_k(n_k(t))\right) & (a) \\ dZ_i = -\frac{\sigma^2 \cdot (-\Delta^+(n_i) + 2Z_i + (\Delta^+(n_i) - 2Z_i)^3)}{2} dt + \sigma \frac{(\Delta^-(n_i))^2 - (\Delta^+(n_i) - 2Z_i)^2}{2(\Delta^-(n_i))} dB & (b) \end{cases}$$ Eq. 10

The true inflow rate $q_{in,ii}(t)$ is given by the following formula if there is enough remaining space in the region:

$$q_{in,ii}(t) = q'_{in,ii}(t) = \begin{cases} q_{MAX}(t), & if \ n_{in,i}^{buf}(t) > 0 \\ q_{in,ii}^{raw}(t), & if \ n_{in,i}^{buf}(t) = 0 \end{cases}$$ Eq. 11

$q'_{in,ii}(t)$ is the potential flow rate into the region. However, the true entering flow rate $q_{in,ii}(t)$ is also limited by the number of vehicles that is already in the region. If the density reaches jam density, then the final entering flow rate would be zero. The number of jammed vehicles is $n_{i,jam}$. Thus, the potential flow rate of vehicles that can enter the region can be approximated by the following formula, if there is enough potential flow rate

$$q_{in,ii}(t) = q'_{in,ii}(t) \frac{(n_{i,jam} - n_i(t))}{\sqrt{M + (n_{i,jam} - n_i(t))^2}} \stackrel{def}{=} q'_{in,ii}(t) \cdot \Psi(n_{i,jam} - n_i(t))$$ Eq. 12

The function $\Psi(x)$ is defined as $\Psi(x) = \frac{x}{\sqrt{M+x^2}}$. $M$ is a positive small enough. The function $\Psi(x)$ maps x to 1 or -1, dependent on the sign of x. The curve of $\Psi(x)$ is given in Figure 6.

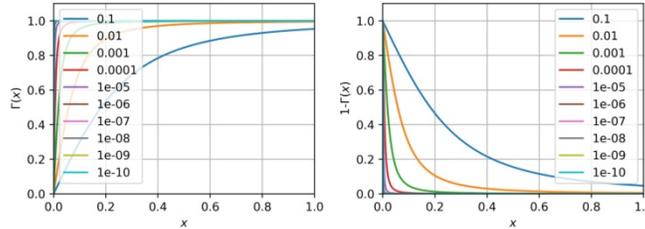

Figure 6. Function $\Psi(x)$ and $1 - \Psi(x)$.

Combining Eq. 11 and Eq. 12, we have the potential flow rate and the final flow rate expressed as follows:

$$q'_{in,ii}(t) = \frac{q_{MAX}(t) \cdot \left(n_{in,i}^{buf}(t)\right)}{\sqrt{\underline{M} + \left(n_{in,i}^{buf}(t)\right)^2}} + q_{in,i}^{raw}(t)\left(1 - \frac{\left(n_{in,i}^{buf}(t)\right)}{\sqrt{\underline{M} + \left(n_{in,i}^{buf}(t)\right)^2}}\right) \quad \text{Eq. 13}$$

$$= q_{MAX}(t) \cdot \Psi\left(n_{in,i}^{buf}(t)\right) + q_{in,i}^{raw}(t)\left(1 - \Psi\left(n_{in,i}^{buf}(t)\right)\right)$$

And final input flow rate $q_{in,ii}(t)$ is calculated as:

$$q_{in,ii}(t) = q_{MAX}(t) \cdot \Psi\left(n_{in,i}^{buf}(t)\right)\Psi\left(n_{i,jam} - n_i(t)\right) \quad \text{Eq. 14}$$
$$+ q_{in,i}^{raw}(t)\left(1 - \Psi\left(n_{in,i}^{buf}(t)\right)\right)\Psi\left(n_{i,jam} - n_i(t)\right)$$
$$\stackrel{\text{def}}{=} \Omega_{in,i}\left(n_{in,i}^{buf}(t), n_i(t)\right)$$

Besides, as we consider the number of vehicles in the buffer area, it should be treated as a state variable:

$$\frac{d\, n_{in,i}^{buf}(t)}{dt} = q_{in,ii}^{raw}(t) - q_{in,ii}(t) = q_{in,i}^{raw}(t) - \Omega_{in,i}\left(n_{in,i}^{buf}(t), n_i(t)\right) \quad \text{Eq. 15}$$

Finally, combining Eq. 10 and Eq. 15, the system state equation for a single region is expressed as follows:

$$\begin{cases} dn_i = \left[\Omega_{in,i}\left(n_{in,i}^{buf}(t), n_i(t)\right) - \left(\frac{n_{ii}(t)}{n_i(t)} + \sum_{k \neq i}\frac{n_{ik}(t)}{n_i(t)}\right)\left(g_{mi}(n_i(t)) + Z(n_i(t))\right) + \sum_{k \neq i}\frac{n_{ki}(t)}{n_i(t)}\left(g_{mi}(n_k(t)) + Z_k(n_k(t))\right)\right]dt & (a) \\ dZ_i = -\frac{\sigma^2 \cdot (-\Delta^+(n_i) + 2Z_i + (\Delta^+(n_i) - 2Z_i)^3)}{2}dt + \sigma\frac{(\Delta^-(n_i))^2 - (\Delta^+(n_i) - 2Z_i)^2}{2(\Delta^-(n_i))}dB & (b) \\ dn_{in,i}^{buf}(t) = \left[q_{in,i}^{raw}(t) - \Omega_{in,i}\left(n_{in,i}^{buf}(t), n_i(t)\right)\right]dt & (c) \end{cases} \quad \text{Eq. 16}$$

There are three state variables: accumulation in region, number of vehicles in the buffer area, and the variation in the rate of exit flow. The true exit flow rate can be calculated as $G(t) = g_{mi}(n_i(t)) + Z(t)$. To express the system state in matrix form, we define the state at Region i as $S_i = [n_i, Z_i, n_{in,i}^{buf}]^T$. Eq. 16 is reformulated as:

$$dS_i = F_i(S_i)dt + L_i(S_i)dB \quad \text{Eq. 17}$$

The matrix $F_i(\cdot)$ and $L_i(\cdot)$ are expressed as:

$$\begin{cases} F_i(S) = \begin{bmatrix} \Omega_{in,i}\left(n_{in,i}^{buf}(t), n_i(t)\right) - \left(\dfrac{n_{ii}(t)}{n_i(t)} + \sum_{k \neq i} \dfrac{n_{ik}(t)}{n_i(t)}\right)\left(g(n_i(t)) + Z(n_i(t))\right) + \sum_{k \neq i} \dfrac{n_{ki}(t)}{n_i(t)}\left(g(n_k(t)) + Z_k(n_k(t))\right) \\ -\dfrac{\sigma^2 \cdot \left(-\Delta^+(n_i) + 2Z_i + (\Delta^+(n_i) - 2Z_i)^3\right)}{2} \\ q_{in,i}^{raw}(t) - \Psi_{in,i}\left(n_{in,i}^{buf}(t), n_i(t)\right) \end{bmatrix} \\ \\ L_i(S) = \begin{bmatrix} 0 \\ \sigma \dfrac{(\Delta^-(n_i))^2 - (\Delta^+(n_i) - 2Z_i)^2}{2(\Delta^-(n_i))} \\ 0 \end{bmatrix} \end{cases}$$

Eq. 18

The boundary condition of Eq. 17 is the raw demand input, i.e. $q_{in,i}^{raw}(t)$ and the initial condition (i.e. $n_i(0)$, $Z_i(0)$ and $n_{in,i}^{buf}(0)$) can be specified, as needed. More than two models of regions can be generalized by extending the system state variables. We omit it here.

In the model development, we do not assume any functional form of the exit MFD (i.e. $g_{mi}(n)$ and the two boundaries in Eq. 16 and Eq. 17). Therefore, it is possible to apply a different functional form and even set different parameters (i.e., $\sigma$) for the regions. We will examine the performances when a different exit MFD form is applied in the Case Study Section.

## 3 Model solution, stability analysis, and calibration method

In this section, we first solve the proposed model in Eq. 16 by means of the Fokker-Planck equation, and then analyze the stability condition, which will be used in the parameter calibration method.

### 3.1 Solution by Fokker-Planck equation

As the proposed model is stochastic, the system state (the number of vehicles in the area, exit flow variation, and in buffer) is a multi-dimensional stochastic process. Its distribution evolves with time, and can solve the famous Fokker-Planck equation. The solution, or marginal distribution of the three variables, is represented by $p(n_i, Z_i, n_{in,i}^{buf}, t)$. For simplicity, we consider the solution of a single region's dynamics. The drift term in the state equation Eq. 17 is simplified to the following formula (the exit flow to other regions can be set to zero):

$$\begin{bmatrix} F_{i1}(S_i) \\ F_{i2}(S_i) \\ F_{i3}(S_i) \end{bmatrix} = \begin{bmatrix} \Omega_{in,i}\left(n_{in,i}^{buf}(t), n_i(t)\right) - \left(g(n_i(t)) + Z(n_i(t))\right) \\ -\dfrac{\sigma^2 \cdot \left(-\Delta^+(n_i) + 2Z_i + (\Delta^+(n_i) - 2Z_i)^3\right)}{2} \\ q_{in,i}^{raw}(t) - \Psi_{in,i}\left(n_{in,i}^{buf}(t), n_i(t)\right) \end{bmatrix}$$

Eq. 19

For easy derivation of the Fokker-Planck equation, we calculate a matrix, $L_i(S_i)QL_i^T(S_i)$, for later use. Q is the variance of the noise and, as the noise is one dimensional, then $Q = \delta t$. Therefore,

$$L_i(S_i)QL_i^T(S_i) = \left[\sigma\frac{0}{\frac{(\Delta^-(n_i))^2 - (\Delta^+(n_i) - 2Z_i)^2}{2(\Delta^-(n_i))}}\right]\delta t\left[\sigma\frac{0}{\frac{(\Delta^-(n_i))^2 - (\Delta^+(n_i) - 2Z_i)^2}{2(\Delta^-(n_i))}}\right]^T \quad \text{Eq. 20}$$

$$= \begin{bmatrix} 0 & 0 & 0 \\ 0 & \delta t \cdot L_{i2}^2(S_i) & 0 \\ 0 & 0 & 0 \end{bmatrix}$$

The Fokker-Planck equation then can be expressed as:

$$\frac{\partial P}{\partial t} = -\left[\frac{\partial(F_{i1} \cdot P)}{\partial n_i} + \frac{\partial(F_{i2} \cdot P)}{\partial Z_i} + \frac{\partial(F_{i3} \cdot P)}{\partial n_{in,i}^{buf}}\right] + \frac{1}{2}\left[\frac{\partial^2(\delta t \cdot L_{i2}^2(S_i) \cdot P)}{\partial Z_i^2}\right] \quad \text{Eq. 21}$$

$$= -\left[\frac{P \cdot \partial(F_{i1})}{\partial n_i} + \frac{F_{i1} \cdot \partial(P)}{\partial n_i} + \frac{P \cdot \partial(F_{i2})}{\partial Z_i} + \frac{F_{i2} \partial(P)}{\partial Z_i} + \frac{F_{i3} \cdot \partial(P)}{\partial n_{in,i}^{buf}}\right.$$

$$\left. + \frac{P \cdot \partial(F_{i3})}{\partial n_{in,i}^{buf}}\right] + \frac{1}{2}\left[\frac{\partial^2(\delta t \cdot L_{i2}^2(S_i) \cdot P)}{\partial Z_i^2}\right]$$

The three partial differentials, i.e., $\frac{\partial(F_{i1})}{\partial n_i}$, $\frac{\partial(F_{i2})}{\partial Z_i}$ and $\frac{\partial(F_{i3})}{\partial n_{in,i}^{buf}}$ can be calculated based on Eq. 19 as:

$$\begin{cases} \frac{\partial(F_{i1})}{\partial n_i} = \frac{\partial\Omega_{in,i}\left(n_{in,i}^{buf}(t), n_i(t)\right)}{dn_i} - (g'(n_i) + Z'(n_i)) \\ \frac{\partial(F_{i2})}{\partial Z_i} = 3\sigma^2(\Delta^+(n_i) - 2Z_i)^2 - \sigma^2 \\ \frac{\partial(F_{i3})}{\partial n_{in,i}^{buf}} = \frac{\partial\Omega_{in,i}\left(n_{in,i}^{buf}(t), n_i(t)\right)}{\partial n_{in,i}^{buf}(t)} \end{cases} \quad \text{Eq. 22}$$

The last term (second-order partial differential in Eq. 21) can be derived from its first partial differential, i.e.

$$\frac{\partial(\delta t \cdot L_{i2}^2(S_i) \cdot P)}{\partial Z_i} = \delta t \cdot P \frac{\partial(L_{i2}^2(S_i))}{\partial Z_i} + L_{i2}^2(S_i)\frac{\partial P}{\partial Z_i} \quad \text{Eq. 23}$$

Then, the second partial differential term is derived as:

$$\frac{\partial^2(\delta t \cdot L_{i2}^2(S_i) \cdot P)}{\partial Z_i^2}$$

Eq. 24

$$= \delta t \cdot \frac{\partial P}{\partial Z_i} \frac{\partial\left(L_{i2}^2(S_i)\right)}{\partial Z_i} + \delta t \cdot P \frac{\partial^2\left(L_{i2}^2(S_i)\right)}{\partial Z_i^2} + \frac{\partial P}{\partial Z_i} \frac{\partial\left(L_{i2}^2(S_i)\right)}{\partial Z_i}$$

$$+ L_{i2}^2(S_i) \frac{\partial^2 P}{\partial Z_i^2}$$

Partial differential term in the first term in RHS of Eq. 24 is calculated as:

$$\frac{\partial\left(L_{i2}^2(S_i)\right)}{\partial Z_i} = \sigma^2 \frac{\partial\left(\left(\frac{\left(\Delta^-(n_i)\right)^2 - \left(\Delta^+(n_i) - 2Z_i\right)^2}{2\left(\Delta^-(n_i)\right)}\right)^2\right)}{\partial Z_i}$$

Eq. 25

$$= \frac{2\left(\Delta^+(n_i) - 2Z_i\right)\left(\left(\Delta^-(n_i)\right)^2 - \left(\Delta^+(n_i) - 2Z_i\right)^2\right) \cdot \sigma^2}{\left(\Delta^-(n_i)\right)^2}$$

Partial differential term in the second term in RHS of Eq. 24 is calculated as

$$\frac{\partial^2\left(L_{i2}^2(S_i)\right)}{\partial Z_i^2} = \frac{2\sigma^2}{\left(\Delta^-(n_i)\right)^2}\left[-2\left(\left(\Delta^-(n_i)\right)^2 - \left(\Delta^+(n_i) - 2Z_i\right)^2\right)\right.$$

Eq. 26

$$\left. + 4\left(\left(\Delta^+(n_i) - 2Z_i\right)\right)^2\right]$$

$$= \frac{2\sigma^2\left[5\left(\left(\Delta^+(n_i) - 2Z_i\right)\right)^2 - 2\left(\Delta^-(n_i)\right)^2\right]}{\left(\Delta^-(n_i)\right)^2}$$

Hence, the last term in Eq. 21 is developed as:

$$\frac{1}{2}\frac{\partial^2(\delta t \cdot L_{i2}^2(S_i) \cdot P)}{\partial Z_i^2}$$

Eq. 27

$$= \frac{1}{2}\left[\delta t \cdot \frac{\partial P}{\partial Z_i} + 1\right]\frac{\partial\left(L_{i2}^2(S_i)\right)}{\partial Z_i} + \frac{1}{2}\delta t \cdot P\frac{\partial^2\left(L_{i2}^2(S_i)\right)}{\partial Z_i^2} + \frac{1}{2}L_{i2}^2(S_i)\frac{\partial^2 P}{\partial Z_i^2}$$

$$= \left[\delta t \cdot \frac{\partial P}{\partial Z_i} + 1\right]\frac{\left(\Delta^+(n_i) - 2Z_i\right)\left(\left(\Delta^-(n_i)\right)^2 - \left(\Delta^+(n_i) - 2Z_i\right)^2\right) \cdot \sigma^2}{\left(\Delta^-(n_i)\right)^2} + \delta t$$

$$\cdot P \frac{\sigma^2\left[5\left(\left(\Delta^+(n_i) - 2Z_i\right)\right)^2 - 2\left(\Delta^-(n_i)\right)^2\right]}{\left(\Delta^-(n_i)\right)^2} + \frac{L_{i2}^2(S_i)}{2}\frac{\partial^2 P}{\partial Z_i^2}$$

Putting the above equations together, we have the solution of the proposed SDE-MFD, which is represented by a time-evolving probability distribution (i.e., $p(n_i, Z_i, n_{in,i}^{buf}, t)$), solves the following Fokker-Planck equation:

$$\frac{\partial \boldsymbol{P}}{\partial t} = -\boldsymbol{P}\left(F_{i1}\frac{\partial \boldsymbol{P}}{\partial n_i} + F_{i2}\frac{\partial \boldsymbol{P}}{\partial Z_i} + F_{i3}\frac{\partial \boldsymbol{P}}{\partial n_{in,i}^{buf}}\right)\left[\frac{\partial \Omega_{in,i}\left(n_{in,i}^{buf}(t), n_i(t)\right)}{dn_i}\right.$$

$$- \left(g'(n_i) + Z'(n_i)\right) + 3\sigma^2(\Delta^+(n_i) - 2Z_i)^2 - \sigma^2$$

$$+ \frac{\partial \Omega_{in,i}\left(n_{in,i}^{buf}(t), n_i(t)\right)}{\partial n_{in,i}^{buf}(t)}\Bigg]$$

$$+ \left[\delta t \cdot \frac{\partial \boldsymbol{P}}{\partial Z_i}\right.$$

$$\left. + 1\right]\frac{(\Delta^+(n_i) - 2Z_i)\left((\Delta^-(n_i))^2 - (\Delta^+(n_i) - 2Z_i)^2\right) \cdot \sigma^2}{(\Delta^-(n_i))^2} + \delta t$$

$$\cdot \boldsymbol{P}\frac{\sigma^2\left[5((\Delta^+(n_i) - 2Z_i))^2 - 2(\Delta^-(n_i))^2\right]}{(\Delta^-(n_i))^2} + \frac{L_{i2}^2(S_i)}{2}\frac{\partial^2 \boldsymbol{P}}{\partial Z_i^2}$$

Eq. 28

### 3.2 Stability analysis

Here we analyze the stability of the stochastic system which is represented by Eq. 16. Just like and ODE, the stability of a system which is described by SDE, have several stability definitions with corresponding physical meanings. We employ the stability definition for stochastic system in Mao (1991). First, we let the drift and diffusion terms be zero and derive the equilibrium points.

$$\begin{cases} \left[\Psi_{in,i}\left(n_{in,i}^{buf}(t), n_i(t)\right) - \left(\frac{n_{ii}(t)}{n_i(t)} + \sum_{k \neq i}\frac{n_{ik}(t)}{n_i(t)}\right)\left(g(n_i(t)) + Z(n_i(t))\right) + \sum_{k \neq i}\frac{n_{ki}(t)}{n_i(t)}\left(g(n_k(t)) + Z_k(n_k(t))\right)\right] = 0 & (a) \\ -\frac{\sigma^2 \cdot (-\Delta^+(n_i) + 2Z_i + (\Delta^+(n_i) - 2Z_i)^3)}{2} = 0 \ ; \sigma\frac{(\Delta^-(n_i))^2 - (\Delta^+(n_i) - 2Z_i)^2}{2(\Delta^-(n_i))} & (b) \\ q_{in,i}^{raw}(t) - \Psi_{in,i}\left(n_{in,i}^{buf}(t), n_i(t)\right) = 0 & (c) \end{cases}$$

Eq. 29

We indicate the equilibrium point as $\left[n_i^{eq}, Z_i^{eq}, n_{in,i}^{buf,eq}\right]^T$. The lyapunov function is constructed as follows:

$$V(n_i, Z_i, n_{in,i}^{buf}) = \sum_i (Z_i - Z_i^{eq})^2 + (n_i - n_i^{eq})^2 + (n_{in,i}^{buf} - n_{in,i}^{buf,eq})^2 \quad \text{Eq. 30}$$

The first-order derivate of $V(n_i, Z_i, n_{in,i}^{buf})$ reads:

$$\frac{\partial V}{\partial n_i} = 2(n_i - n_i^{eq}); \frac{\partial V}{\partial Z_i} = 2(Z_i - Z_i^{eq}); \frac{\partial V}{\partial n_{in,i}^{buf}} = 2(n_{in,i}^{buf} - n_{in,i}^{buf,eq}) \quad \text{Eq. 31}$$

The second derivate of $V(n_i, Z_i, n_{in,i}^{buf})$ reads:

$$\begin{cases} \frac{\partial^2 V}{\partial Z_i^2} = 2; \frac{\partial^2 V}{\partial n_i^2} = 2; \frac{\partial^2 V}{\partial (n_{in,i}^{buf})^2} = 2; \\ \frac{\partial^2 V}{\partial \# \partial *} = 0, \forall \#, * \in \{n_i, Z_i, n_{in,i}^{buf}\}, \# \neq * \end{cases} \quad \text{Eq. 32}$$

According to the Ito rule, we can derive the expression of $V(n_i, Z_i, n_{in,i}^{buf})$ as a stochastic differential equation. Given the state equation in Eq..16, according to the Ito rule, we write the equation for $V(n_i, Z_i, n_{in,i}^{buf})$ as:

$$dV = LV(n_i, Z_i, n_{in,i}^{buf}) \cdot dt + \left[\frac{\partial V}{\partial n_i}, \frac{\partial V}{\partial Z_i}, \frac{\partial V}{\partial n_{in,i}^{buf}}\right] \cdot L_i(\cdot) \cdot dB \quad \text{Eq. 33}$$

$LV(n_i, Z_i, n_{in,i}^{buf})$ in the above equation is obtained via:

$$LV(\cdot) = \left[\frac{\partial V}{\partial n_i}, \frac{\partial V}{\partial Z_i}, \frac{\partial V}{\partial n_{in,i}^{buf}}\right] F_i\left(n_i, Z_i, n_{in,i}^{buf}\right) + \frac{1}{2} trace\left(L_i^T \begin{bmatrix} 2 & \cdots & 0 \\ \vdots & \ddots & \vdots \\ 0 & \cdots & 2 \end{bmatrix} L_i\right) \quad \text{Eq. 34}$$

$$= 2(n_i - n_i^{eq})\left[\Psi_{in,i}\left(n_{in,i}^{buf}(t), n_i(t)\right) - \left(\frac{n_{ii}(t)}{n_i(t)} + \sum_{k \neq i}\frac{n_{ik}(t)}{n_i(t)}\right)\left(g(n_i(t)) + z(n_i(t))\right)\right.$$

$$\left. + \sum_{k \neq i}\frac{n_{ki}(t)}{n_i(t)}\left(g(n_k(t)) + Z_k(n_k(t))\right)\right]$$

$$- 2(Z_i - Z_i^{eq})\frac{\sigma^2 \cdot (-\Delta^+(n_i) + 2Z_i + (\Delta^+(n_i) - 2Z_i)^3)}{2}$$

$$+ 2\left(n_{in,i}^{buf} - n_{in,i}^{buf,eq}\right)\left[q_{in,i}^{raw}(t) - \Psi_{in,i}\left(n_{in,i}^{buf}(t), n_i(t)\right)\right]$$

$$+ \sigma^2\left(\frac{\left(\Delta^-(n_i)\right)^2 - (\Delta^+(n_i) - 2Z_i)^2}{2(\Delta^-(n_i))}\right)^2$$

The stability condition of the stochastic differential equation requires that the function $LV(\cdot)$ be non-positive, i.e.

$$LV(\cdot) \leq 0 \quad \text{Eq. 35}$$

Let $LV_1(n_i, Z_i, n_{in,i}^{buf})$ and $LV_2(n_i, Z_i, n_{in,i}^{buf})$ be:

$$LV_1(\cdot,\cdot,\cdot) = (-\Delta^+(n_i) + 2Z_i + (\Delta^+(n_i) - 2Z_i)^3)(Z_i - Z_i^{eq}) - \left(\frac{\left(\Delta^-(n_i)\right)^2 - (\Delta^+(n_i) - 2Z_i)^2}{2(\Delta^-(n_i))}\right)^2 \quad \text{Eq. 36}$$

And

$$LV_2(\cdot,\cdot,\cdot) = 2(n_i - n_i^{eq})\left[\Psi_{in,i}\left(n_{in,i}^{buf}(t), n_i(t)\right) - \left(\frac{n_{ii}(t)}{n_i(t)} + \sum_{k \neq i}\frac{n_{ik}(t)}{n_i(t)}\right)\left(g(n_i(t)) + Z(n_i(t))\right)\right. \quad \text{Eq. 37}$$

$$\left. + \sum_{k \neq i}\frac{n_{ki}(t)}{n_i(t)}\left(g(n_k(t)) + Z_k(n_k(t))\right)\right]$$

$$+ 2\left(n_{in,i}^{buf} - n_{in,i}^{buf,eq}\right)\left[q_{in,i}^{raw}(t) - \Psi_{in,i}\left(n_{in,i}^{buf}(t), n_i(t)\right)\right]$$

respectively. Then the stability condition is satisfied, if the following condition holds:

$$\sigma^2 \leq \frac{LV_2(n_i, Z_i, n_{in,i}^{buf})}{LV_1(n_i, Z_i, n_{in,i}^{buf})}, \forall n_i, Z_i, n_{in,i}^{buf} \qquad \text{Eq. 38}$$

### 3.3 Calibration method

There is one parameter in the model (i.e., σ). The parameter describes the randomn level of the target region. Given the observation, we proceed to develop a method for the calibration of parameter σ. The observation is give by a series of $n_i(k)$ for Region i for k time step. Hence, the given observation can be expressed as $O = \{o_i(k)\} = \{(n_i(k))\}$. The likelihood of the observation then is expressed as:

$$L(O|\sigma) = p_\sigma(n_0(1)) \prod_{k=1}^{N} p_\sigma(n_i(k)|n_{i-1}(k)) \qquad \text{Eq. 39}$$

Thus, the calibration can be implemented as:

$$\sigma^* = \max_\sigma L(O|\sigma) \qquad \text{Eq. 40}$$

Then, the key of the calibration is the calculation of the conditional probability $p_\sigma(n_i(k)|n_{i-1}(k))$. To derive this conditional probability, we use Eq. 10 to illustrate the calibration process. We discretize the system using the Euler method and we have the discretization form of Eq. 10:

$$n_i(k+1) - n_i(k) = \delta t$$
$$\cdot \left[ q_{in,ii}(k) - \left( \frac{n_{ii}(k)}{n_i(k)} + \sum_{m \neq i} \frac{n_{im}(k)}{n_i(k)} \right) \left( g_i(n_i(k)) + Z_i(n_i(k)) \right) \right.$$
$$\left. + \sum_{m \neq i} \frac{n_{mi}(k)}{n_i(k)} \left( g_m(n_m(k)) + Z_m(n_m(k)) \right) \right] \qquad \text{Eq. 41}$$

$$\xrightarrow{\text{consider single region}} \delta t \cdot \left[ q_{in,ii}(k) - \left( g_i(n_i(k)) + Z_i(n_i(k)) \right) \right]$$

And

$$Z_i(k+1) - Z_i(k) = -\delta t \frac{\sigma^2 \cdot \left( -\Delta^+(n_i(k)) + 2Z_i + \left( \Delta^+(n_i(k)) - 2Z_i(k) \right)^3 \right)}{2}$$
$$+ \sigma \frac{\left( \Delta^-(n_i(k)) \right)^2 - \left( \Delta^+(n_i(k)) - 2Z_i(k) \right)^2}{2 \left( \Delta^-(n_i(k)) \right)} \delta B(k) \qquad \text{Eq. 42}$$

In the above discretization form, $\delta B(k)$ is Gaussian distributed. Its mean is zero and variance is δt, which is the time step of the observation. From Eq. 42, it can be concluded that given $n_i(k-1)$ and $z_i(k-1)$, $z_i(k)$ is Gaussian distributed and its mean and variance are given by:

$$\begin{cases} \mu_{Z_i(k)} = -\delta t \dfrac{\sigma^2 \cdot \left(-\Delta^+\big(n_i(k-1)\big) + 2Z_i(k-1) + \big(\Delta^+\big(n_i(k-1)\big) - 2Z_i(k-1)\big)^3\right)}{2} & \text{Eq. 43} \\[2ex] \sigma^2_{Z_i(k)} = \left(\sigma \dfrac{\big(\Delta^-\big(n_i(k-1)\big)\big)^2 - \big(\Delta^+\big(n_i(k-1)\big) - 2Z_i(k-1)\big)^2}{2\big(\Delta^-\big(n_i(k-1)\big)\big)}\right)^2 \delta t & \end{cases}$$

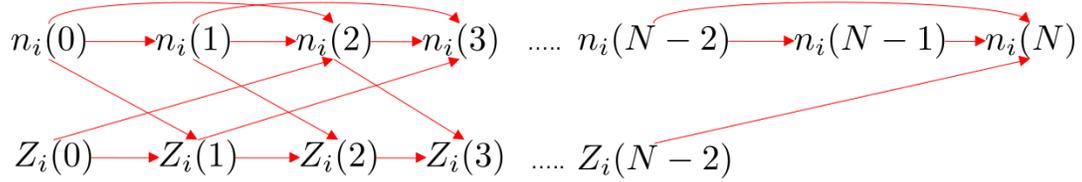

Figure 7. Inter-dependence among the variables.

From Eq. 41, it is concluded that given $n_i(k-1)$ and $z_i(k-1)$, $n_i(k)$ can be calculated. Because $z_i(k-1)$ depends on $z_i(k-2)$, then given $z_i(k-2)$ and $n_i(k-1)$, $n_i(k)$ is a Gaussian distributed variable (as can be seen from Eq. 41), and its mean can be expressed as:

$$\mu_{n_i(k)} = n_i(k-1) + \delta t \cdot \left[ q_{in,ii}(k-1) - \left( g_i\big(n_i(k-1)\big) \right. \right.$$
$$\left. \left. - \delta t \dfrac{\sigma^2 \cdot \left(-\Delta^+\big(n_i(k-2)\big) + 2Z_i(k-2) + \big(\Delta^+\big(n_i(k-2)\big) - 2Z_i(k-2)\big)^3\right)}{2} \right) \right]$$

Eq. 44

And variance is:

$$\sigma^2_{n_i(k)} = \delta t^2 \cdot \sigma^2_{Z_i(k-1)} = \left(\sigma \dfrac{\big(\Delta^-\big(n_i(k-2)\big)\big)^2 - \big(\Delta^+\big(n_i(k-2)\big) - 2Z_i(k-2)\big)^2}{2\big(\Delta^-\big(n_i(k-2)\big)\big)}\right)^2 \delta t^3 \quad \text{Eq. 45}$$

The inter-dependence relationships among variables are given in Figure 7. Hence, the likelihood is expressed as:

$$L(O|\sigma) = p_\sigma(n_i(0)) \prod_{k=1}^{N} p_\sigma(n_i(k)|n_{i-1}(k))  \quad \text{Eq. 46}$$

$$= \int \{p_\sigma(n_i(0))p_\sigma(n_i(1)|n_i(0)) \cdot p_\sigma(Z_i(0))$$
$$\cdot [p_\sigma(Z_i(1)|n_i(0), Z_i(0)) \cdot p(n_i(2)|n_i(1), Z_i(0))]$$
$$\cdot [p_\sigma(Z_i(2)|n_i(1), Z_i(1))$$
$$\cdot p(n_i(3)|n_i(2), Z_i(1))] \cdots [p_\sigma(Z_i(N-1)|n_i(N-2), Z_i(N-2))$$
$$\cdot p(n_i(N)|n_i(N-1), Z_i(N-2))]\} d\, Z_i(0) dZ_i(1) \ldots dZ_i(N-2)$$

In the above formula, both $p_\sigma(Z_i(j)|n_i(j-1), Z_i(j-1))$ and $p(n_i(j)|n_i(j-1), Z_i(j-2))$ are Gaussian distributed. The likelihood $L(O|\sigma)$ can be evaluated. The beginning three terms within the integrand (i.e., $p_\sigma(n_i(0)), p_\sigma(n_i(1)|n_i(0))$ and $p_\sigma(Z_i(0))$ cannot be expressed explicitly. However, when the observation series is long enough, the three terms contribute little to the final probability, and they can be neglected (Saarkk and Solin, 2019). Finally, $\sigma^*$ can be obtained via:

$$\sigma^* = \underset{\sigma}{\mathrm{argmax}}\, L(O|\sigma)\, L(O|\sigma) \quad \text{Eq. 47}$$

$$\approx \underset{\sigma}{\mathrm{argmax}} \int \left\{ \prod_{i=2}^{N} [p_\sigma(Z_i(i-1)|n_i(i-2), Z_i(i-2)) \cdot p(n_i(i)|n_i(i-1), Z_i(i-2))] \right\} d\, Z_i(0) dZ_i(1) \ldots dZ_i(N-2)$$

Unfortunately, the calculations need to resort to Monte Carla integration methods due to the integration. The optimization can be implemented by the PSO algorithm. As PSO is a famous optimization method, details are not presented here.

## 4 Cases study

In this section, we apply our model using calibrated MFD results (polynomial type) from current literature. Then, we examine the performance of the model by applying it to a different but also a popular exit-MFD family (exponential type), which is fitted by the data from software simulation. In both cases, we add certain stochasticity to the system dynamics and examine the sensitivities, the system evolution path, the marginal distribution, and etc.

## 4.1 Single region case

We use the polynomial function fitted in Ding et al. (2017) to investigate the system dynamics of the proposed model. The functional form of the MFD is expressed as $g(n_i(t)) = a_i n_i(t)^3 + b_i n_i(t)^3 + c_i n_i(t)^3$. The parameters include $a_i$, $b_i$ and $c_i$. The parameters for the lower and upper boundaries of this exit-MFD are given in Ding et al. (2017). They are $a_i = 3.298e - 11$, $b_i = -7.37423e-7$ and $c_i = 4.52e-3$. The upper boundary, lower boundary and average are plotted in Figure 8-left subplot. The difference between the upper and lower boundary is given in Figure 8-right subplot.

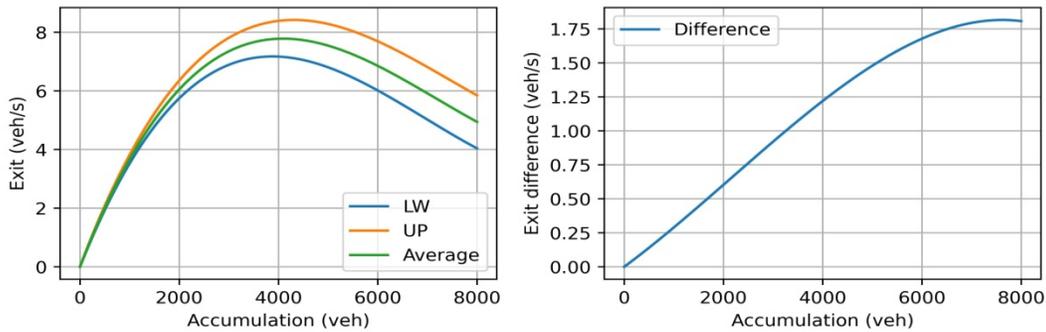

Figure 8. The calibrated MFD.

### 4.1.1 System dynamics

We use the Euler discretization scheme (i.e., Eq. 41 and Eq. 42). To create the congestion evolution and the recovery path, we set a parabolic demand curve, as given in Figure 9-left subplot. There is a peak hour flow, and a constant input demand that follows it. The system states space (there are 1,000 paths, and each path corresponds to one simulation of 5,000 sec) is presented in Figure 9-right subplot. It shows that the system dynamics traverse a certain state-space, rather than a deterministic functional curve. Also, Figure 9-right subplot identifies that the system returns in the lower branch. This is a hysteresis phenomenon, and we will explore it later.

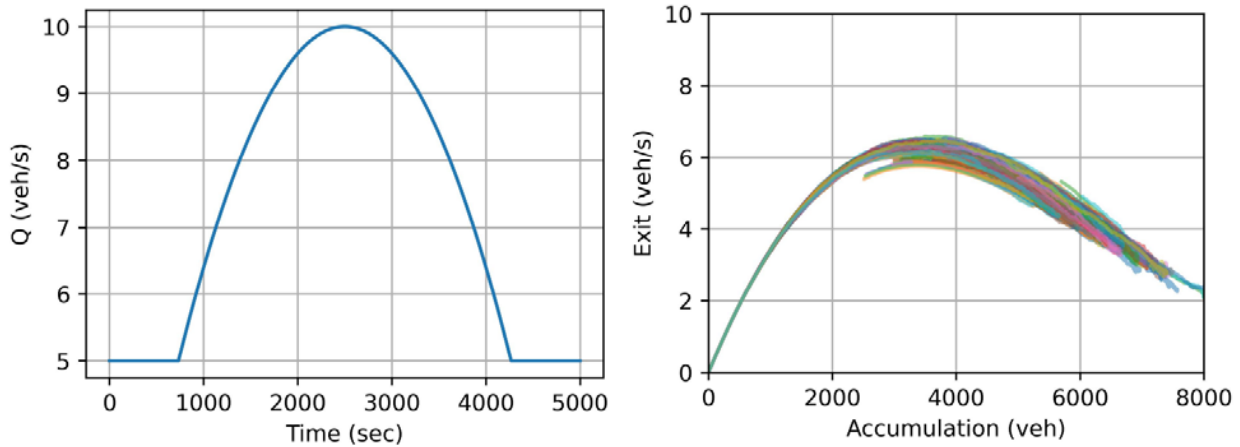

Figure 9. The demand curve and system states from Monte Carlo simulation, σ = 0.04.

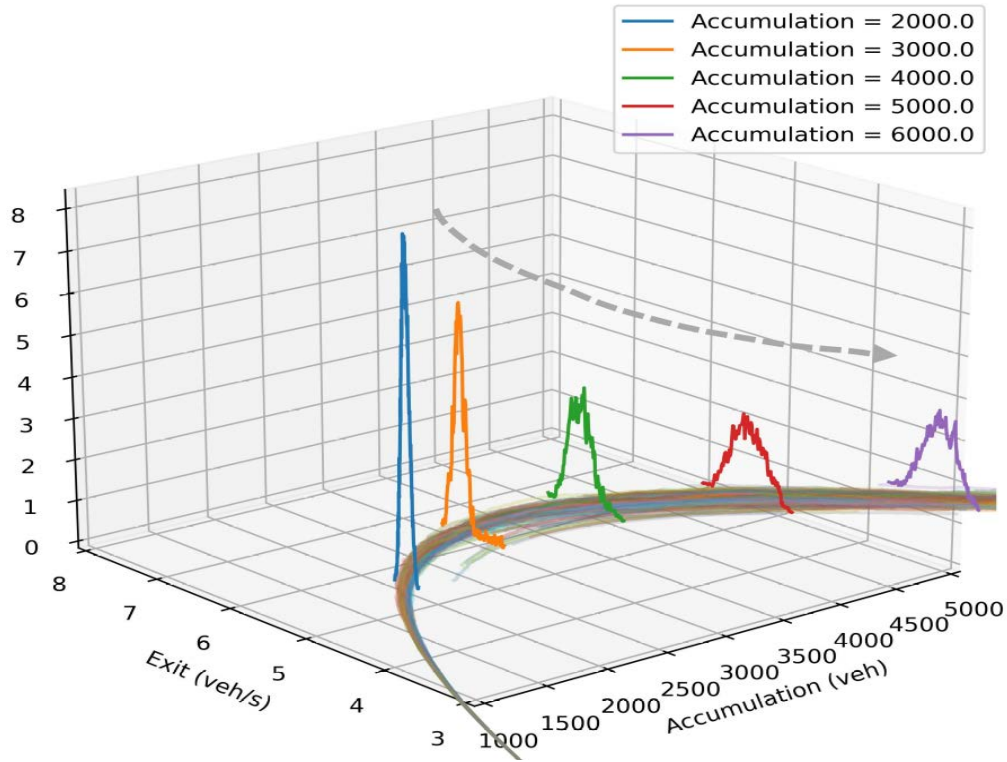

Figure 10. The distribution of the exit flow rate under certain accumulations.

Given the same accumulation, the exit flow follows a certain distribution as the system is stochastic. The distributions under given accumulations are given in Figure 10. The selected accumulations are from 2,000 vehicles to 6,000 vehicles. The distribution curve is normalized, i.e., the integration of the curve is 1. It shows that the distribution varies for different accumulations. As an accumulation increases, a wider spread is observed, as indicated by the dashed curve in Figure 10. Therefore, the model can reproduce the density-dependent scattered phenomena of the macroscopic system dynamics.

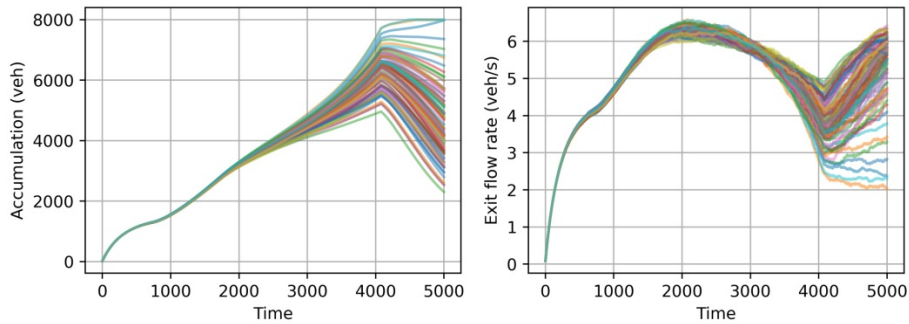

Figure 11. System state paths: (left) accumulation and (right) exit flow.

To better understand the system evolution, we analyze the temporal trajectories. Figure 11 presents the trajectories for accumulation and the exit flow. The accumulation reaches its maximum in about 4,000 sec, although the demand drops at 2,500 sec. Critical density (which corresponds to the maximal exit flow) is about 3,000 vehicles. The accumulation exceeds this critical accumulation at about 2,000 sec and, thus, the exit flow decreases continuously after 2,000 sec (this can be identified in the Figure 11-right subplot). The system thus enters a jam state after 2,000 sec and recovers in about 4,000 sec. However, there are also some possibilities (some paths in Figure 11-left subplot lead to a higher accumulation, even after 4,000 sec) as the jam continues to deteriorate. This can be interpreted as the gridlock process due to random factors such as chain spillover.

Some selected moments of the marginal distribution of the system states (the accumulation and the exit flow rate) are given in Figure 12-left subplot for accumulation, and in the right subplot for exit flow variation. At first, as the exit flow variation is trivial, the system states exhibit certain concentrations. However, after 2,000 sec, the exit flow fluctuation displays considerable diversity (Figure 12-right subplot). The system states then spread over a certain interval.

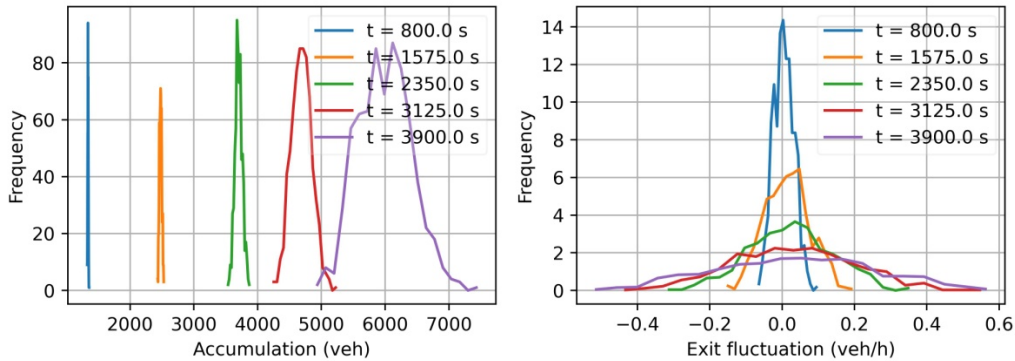

Figure 12. Marginal distribution of accumulation (left) and the exit flow rate (right).

### 4.1.2 Reproducing the hysteresis phenomena

The hysteresis phenomenon refers to the scenario where the unloading path is different from the loading path of the system states. This is proven by the data (Geroliminis and Sun, 2011) and the theoretical model (Gayah and Daganzo, 2011). From the Figure 9-right subplot, the hysteresis can be clearly identified, as there is a lower branch that represents the recovering process. Figure 13 presents two system evolution paths from the simulation under the demand given in Figure 9. We set the accumulation interval as [3,000 vehicles, 6,000 vehicles]. The system first enters this interval and then recovers from the jammed state. As the system is stochastic, the exiting path is different from the entering path in each realization (i.e., simulation) and the exiting path also differs for each simulation. Two paths are selected as representative. For path 1 in Figure 13, the recovering trajectory almost overlaps the entering trajectory, which indicates that the exit flow does not deteriorate. While for path 2, the recovering trajectory is lower than the entering path. The difference here is referred to as the hysteresis difference or decrease. The decrease indicates a loss in capacity. One objective of perimeter control is to prevent this from happening.

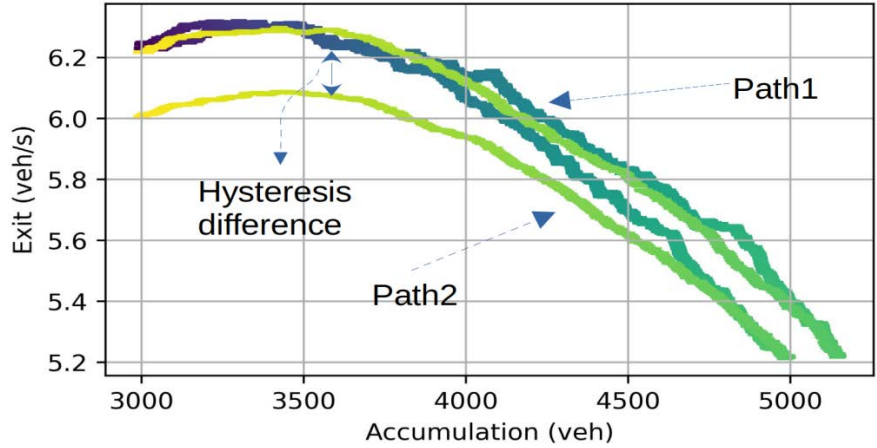

Figure 13. The hysteresis of the simulated paths.

As the system is stochastic, the hysteresis decrease is also stochastic. For certain accumulations, we record the exit flow rate when the path traverses this accumulation (i.e., a jam occurs), and then record the exit flow rate again when the path recovers (i.e., the jam disperses), and then calculate the hysteresis decrease. The average of the hysteresis can be calculated then for all simulated paths. The hysteresis curve is given in Figure 14 and shows that hysteresis exists over a wide range of accumulations and increases with accumulation.

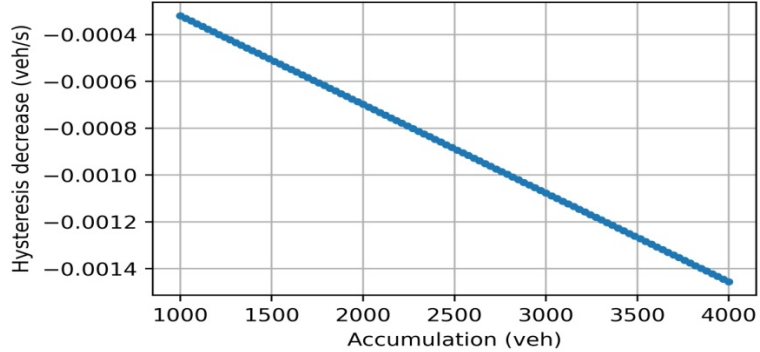

Figure 14. Average hysteresis decreases for all simulated paths.

### 4.2 System dynamics of a different exit-MFD form

The proposed model (i.e., Eq. 16) does not assume any prior MFD type. Except for the polynomial MFD examined in the above sections, there are many other functional forms. In this section, we simulate the functional form in Ramezani, Haddad, and Geroliminis (2015). The functional form is given by $G(n) = p_1 n^{p_2} e^{-\frac{n^{p_2}}{n_{crt}^{p_2}}}$. It has three parameters: $n_{crt}$, $p_1$ and $p_2$. We use a network to simulate the accumulation and the exit flow rate and then fit the parameters. The simulated network consists of 16 intersections (4*4 gird). The resulting accumulation-exit scatter plot is given in Figure 15. The fitted upper boundary curve parameters are: $p_1 = 4.7093 * 10^{-2}$, $p_2 = 1.4137$, $n_{crt} = 1408.4875$, and that for the lower boundary curve are $p_1 = 1.5874 * 10^{-3}$, $p_2 = 1.8538$, $n_{crt} = 1502.2319$. The above two curves are also plotted in Figure 15. According to the scatter, the jam accumulation of the network is set to 8,000 vehicles.

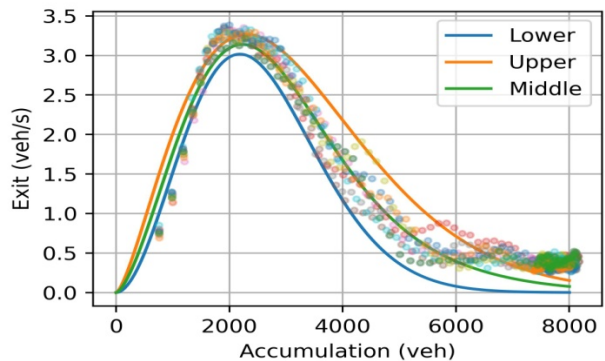

Figure 15. Simulation network and the resulting exit function (right).

The system demand curve and the resulting state trajectories are given in Figure 16. Similar to Figure 9-left subplot, we set a peak hour flow. The time horizon is 5,000 sec. The accumulation increases along with the demand input. After 1,000 sec (the demand suddenly increases according to the parabolic curve), and the accumulation also increases. At about 2,500 sec, the system reaches its critical accumulation (about 2,500 vehicles). After that, the system state diverges: in Figure 9-middle subplot and, after 3,000 sec, some trajectories lead to a bad state, i.e., accumulation continues to increase. Other trajectories lead to the normal state (where the accumulation decreases). The former corresponds to the stochastic gridlock process, where the exit flow drops (in Figure 9-right subplot, however, exit flow of some trajectories continues to decrease).

The system state space and the marginal distribution of exit flow are given in Figure 17. As already discovered in Figure 16-b and c, some trajectories return back, even when demand decreases, while some trajectories continue to become more deteriorated. The state of the-space visited in Figure 17-a is unevenly distributed at the right branch of the exit MFD. Also, the marginal distribution of the exit flow for different moments can be identified in the Figure 17-b subplot.

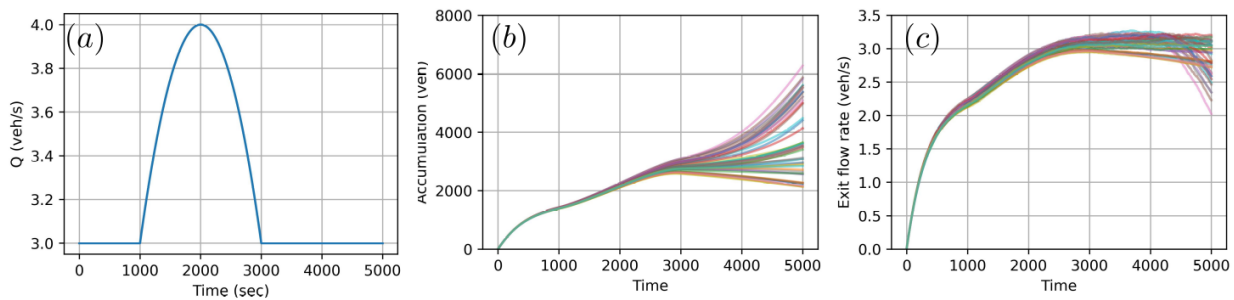

Figure 16. Demand curve and the system path.

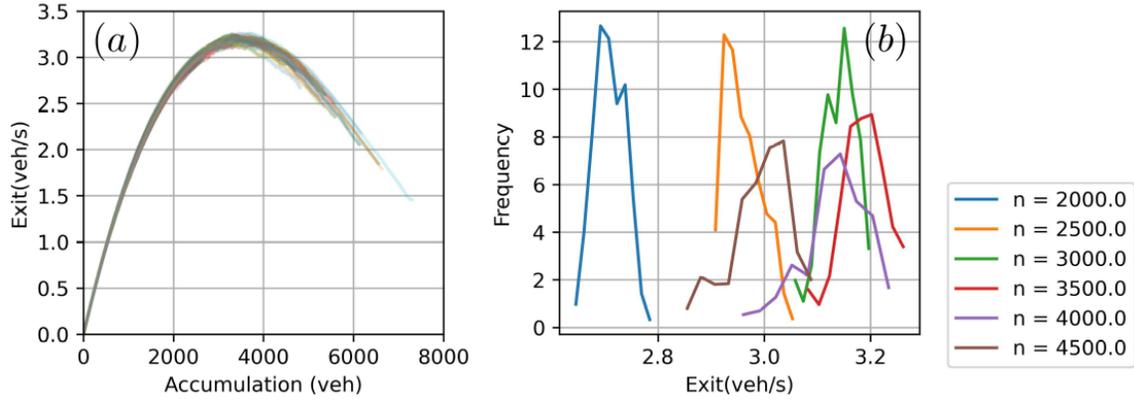

Figure 17. Exit flow distribution given accumulation.

### 4.3 Two regions case

In this section, we investigate the model capabilities by simulating two neighboring regions. More than two regions are just natural generalizations. A two-region framework is widely used in literature to implement perimeter control (Li, Mohajerpoor, and Ramezani 2021; Haddad, 2017; Geroliminis, Haddad, and Ramezani, 2013) or traffic flow dynamics analysis (Sirmatel and Geroliminis, 2021). To simulate the system dynamics of different demand inputs, we set two different peak hours for the two regions. The demand pattern and the resulting MFD are given in Figure 18. In the Figure 18-a subplot, Region 1 experiences the congestion first and, then it is followed by the peak hour of Region 2. The transfer ratio from Region 1 to 2 is 0.7 (i.e., 70% of the demand generated in Region 1 travel to Region 2), and the transfer ratio from Region 2 to 1 is 0.5. As most of the demand of Region 1 travels to Region 2, Region 2 experiences a higher accumulation (nearly 8,000 vehicles) than Region 1 (about 7,000 vehicles).

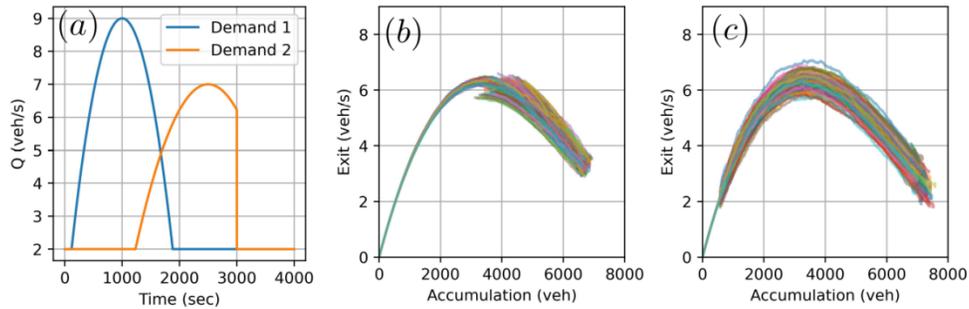

Figure 18. Demand (a), MFD of Region 1 (b) and MFD of Region 2 (c).

Part of the states of trajectories for the two regions are given in Figure 19. Figures 19-a and b are for Region 1 and Figures 19-c and d are for Region 2. The simulation time horizon is 4,000 sec. For Region 1, the demand drops after 1,000 sec (as shown in Figure 18-a subplot). Its accumulation continues to increase until about 1,800 sec. As the accumulation exceeds 4,000 vehicles (which is the congestion state, as in Figure 18-b subplot), Region 1 experiences a jam state within about 1,000 sec to 3,000 sec. From about 1,800 sec, the exit flow of Region 1 increases (in Figure 19-b), which indicates a recovery process. For Region 2, the accumulation first increases (from 0 sec to about 500 sec, as in Figure 19-c) and, then, remains relatively stable (from 500 sec to about 1,000 sec, as in Figure 19-c) and then increases again. The first increase (from 0 sec to about 500 sec) corresponds to the demand from Region 1, and the second increase (after 1,000 sec) corresponds to the demand peak in Region 2 itself. The patterns in Figure 19-d can be similarly identified.

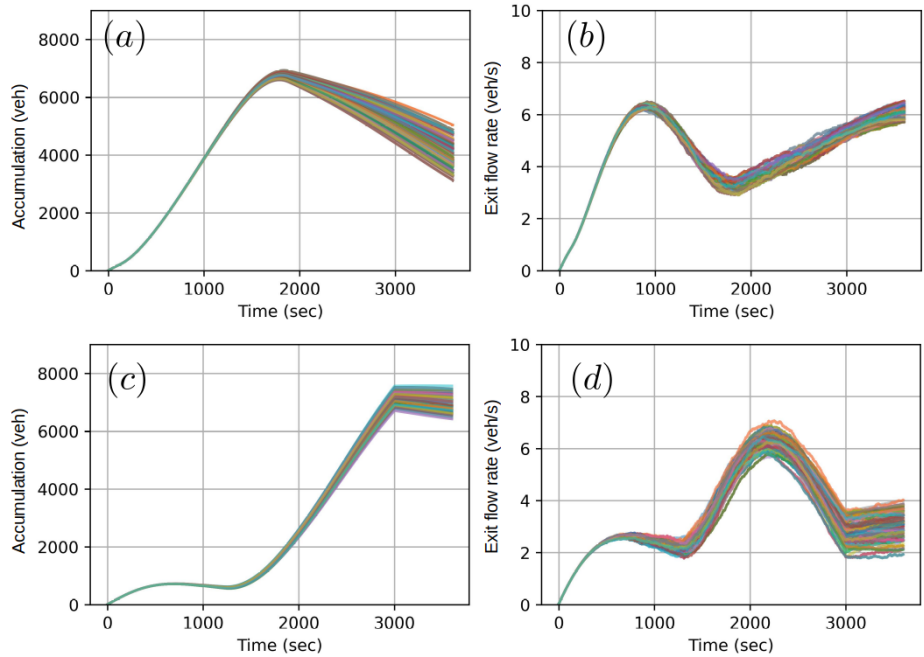

Figure 19. State of system trajectories in Region 1, (a) and (b) and in Region 2:, (c) and (d):

As the system evolution is stochastic, some correlation structure exists between the two regions. The joint distributions (which are displayed as a two-dimensional heat map) are shown in Figure 20. Figure 20-a is for accumulation; while Figure 20-b is for exit flow variation, and Figure 20-c is for exit flow. As we assume that the exit variations for the two regions are independent, the correlation between exit flow variations appears weak in Figure 20-b plot. In contrast, the correlation structure pattern between accumulations and the exit flow rate can be clearly identified. Joint marginal distributions of the states (for instance, the accumulations of the two regions) can be obtained by simulation (which is omitted here).

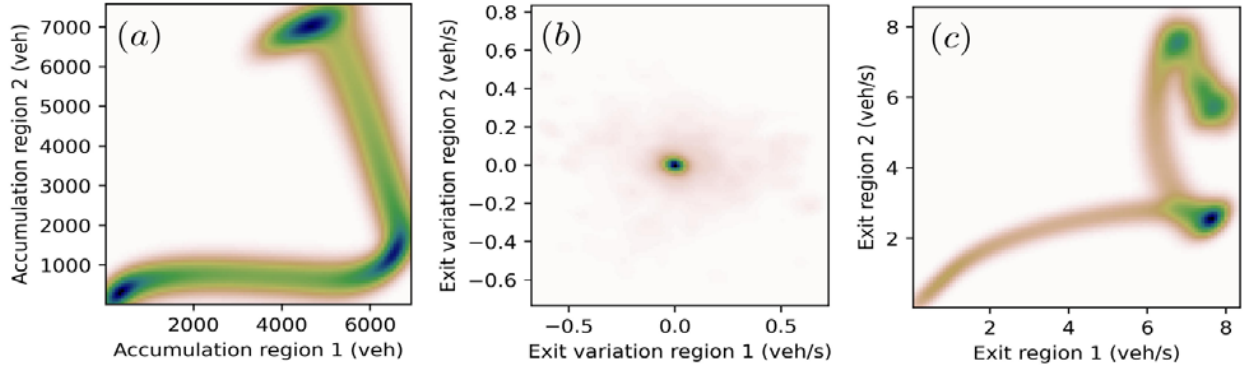

Figure 20. Correlation between accumulations (left), exit flow variation (middle), and exit flow (right).

### 4.4 Sensitivity simulation

In the cases studied above, we assumed that the expectation (i.e., $g_{mi}(n)$ in Figure 4) is located at the middle between the upper and lower boundaries. In reality, this may not be the case when the exit distributes asymmetrically, given the same accumulation. We use a parameter η to express this skew. The expectation is calculated as $g_{mi}(n) = g_{lw}(n) + \eta\left(g_{up}(n) - g_{lw}(n)\right)$. When η = 0.5, the resulting expectation is located at the middle of $g_{up}(n)$ and $g_{lw}(n)$. A greater η indicates that the expectation is nearer to $g_{up}(n)$. We select η = 0.8 and η = 0.2 to examine the difference. The results (MFD and the exit distribution) are given in Figure 21 and Figure 22. By comparing the distributions in Figure 21 and Figure 22, it is found that, when η = 0.8, the skew of the exit flow distribution is negative (i.e., the distribution concentrates on the right), and the system states recover more easily (the maximal accumulation is lower than that when η = 0.2). When η = 0.2, the skew is positive. Therefore, the proposed model can adjust the definition of $g_{mi}(n)$ so that the desired skew can be obtained.

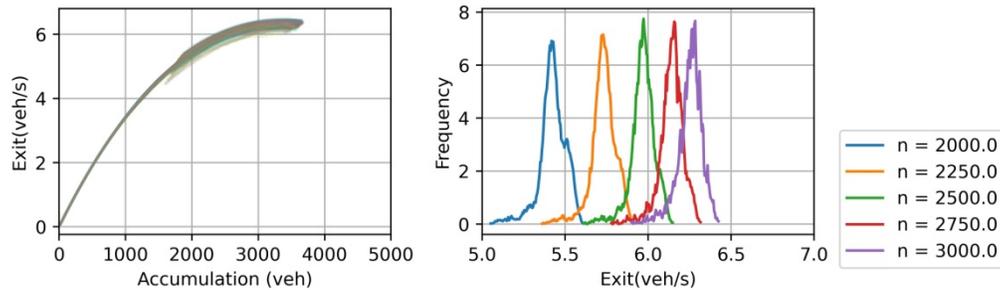

Figure 21. η = 0.8, left subplot-MFD; right subplot-the exit distribution.

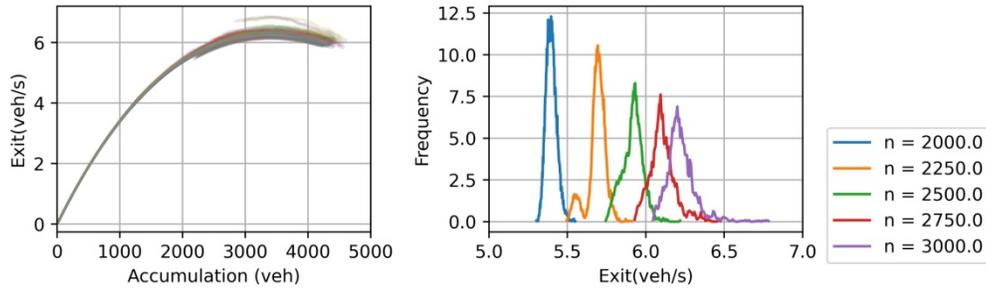

Figure 22. η = 0.2, left subplot-MFD; right subplot-the exit distribution.

The parameter σ indicates the noise level of the system, and reflects the total randomness of the demand, supply, and operational state. The two noise levels, σ = 0.007 and σ = 0.002, that we chose to demonstrate the results are shown in Figure 23. The MFD is similar. When σ = 0.002, the MFD shape is more concentrated. The noise level can be clearly distinguished by the exit flow distribution, under the same accumulation. When distributions σ = 0.007, the states are spread much wider than those of σ = 0.002. Thus, the proposed model can be adjusted for different random levels.

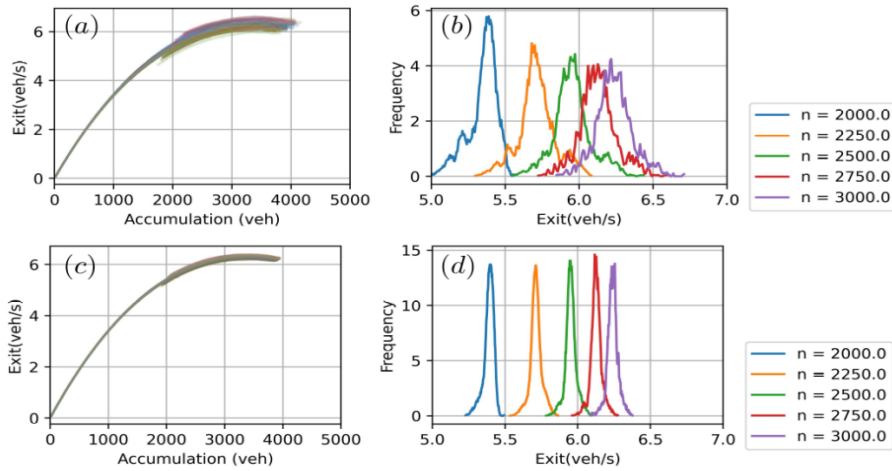

Figure 23. Different nosie levels; (a) and (b): σ = 0.007; (c) and (d): σ = 0.002

## 5 Conclusion and future work

As MFD is getting more and more attractive in traffic flow-related operational applications, the analytical framework of MFD becomes increasingly significant. As the network traffic flow state is random, the stochastic viewpoint of the MFD is necessary to deal with the uncertainty in network traffic flow operations. Yet, current literature cannot afford such analytical tools to deal with stochastic variations and generate the desired aggregated variables distributions. This manuscript proposes a stochastic MFD model, which admits a density-dependent exit flow variation. The model is constructed from the basic Brownian motion model. The stability of the model is analyzed, and the parameter calibration procedure is derived. Validation proves that the model can generate a distribution of aggregated variables. Either stochastic gridlock or hysteresis phenomenon can be reproduced by the model.

Future research directions may include the following: 1) for a large city area, a feasible strategy is to divide the area into a series of sub-zones. Each sub-zone obeys the stochastic traffic flow operational law sketched in this manuscript. The coordination control of multi-stochastic-sub-zones needs to be considered. Such problems can be cast into a stochastic control problem and can be solved using current methods; 2) generally, the data source may not be complete. The methods that estimate, infer, or filter desired variables (e.g., the input flow rate), given partial observables, can be constructed based on the proposed stochastic model; 3) The application of the proposed model to perimeter control, congestion pricing, etc. can be considered. This will be the next step for this work.

### Acknowledgement

Project supported by the Key Research and Development Program of China (2021YFE0194400) and the National Natural Science Foundation of China (No. 52131202); The ministry of education in China project of humanities and social science (21YJCZH116) and Zhejiang province public welfare scientific research project(LGF22E080007).